\documentclass[namedreferences]{SolarPhysics}
\usepackage[optionalrh]{spr-sola-addons}

\usepackage{graphicx} 
\begin{document}
\begin{article}
\begin{opening}
\title{Including stereoscopic information
in the reconstruction of coronal magnetic fields}

\author{T. \surname{Wiegelmann}\email{tw@mcs.st-and.ac.uk}}
\author{T. \surname{Neukirch}} 
\institute{School of Mathematics and Statistics,
  University of St. Andrews,
  St. Andrews, KY16 9SS,
  United Kingdom}

\date{Solar Physics, v. 208, Issue 2, p. 233-251 (2002)}

\runningtitle{Stereoscopic information in reconstruction}
\runningauthor{Wiegelmann and Neukirch}

\begin{abstract}
We present a method to include stereoscopic information about the
three dimensional structure of flux tubes into the reconstruction
of the coronal magnetic field. Due to the low plasma beta in the
corona we can assume a force free magnetic field, with the current
density parallel to the magnetic field lines. Here we use linear
force free fields for simplicity. The method uses the line of
sight magnetic field on the photosphere as observational input.
The value of $\alpha$ is determined iteratively by comparing the
reconstructed magnetic field with the observed structures. The
final configuration is the optimal linear force solution
constrained by both the photospheric magnetogram and the observed
plasma structures. As an example we apply our method to SOHO
MDI/EIT data of an active region. In the future it is planned to
apply the method to analyse data from the SECCHI instrument aboard
the STEREO mission.
\end{abstract}


\end{opening}
\section{Introduction}

Due to the low average plasma $\beta$ the structure of the corona
is determined by the coronal magnetic field. Knowledge of the
structure of the coronal magnetic field is therefore of prime
importance to understand the physical processes in the solar
corona.

At the present time there is no general method available which
allows the direct and accurate measurement of the magnetic field
at an arbitrary point in the corona, although some progress has
been made using radio observations above active regions (e.g.
\opencite{golub97}). We therefore have to extrapolate the coronal
magnetic field from measurements taken at photospheric or
chromospheric level.

If we want to do this we have to make assumptions about the
current density in the corona. The low average plasma $\beta$
allows us to assume that to lowest order the magnetic field is
force-free, i.e. the current density is aligned with the magnetic
field. With a few exceptions (e.g. \opencite{zhao93};
\opencite{zhao94}; \opencite{petrie00}; \opencite{zhao00};
\opencite{rudenko01b}), most of the extrapolation and
reconstruction methods proposed so far are based on this
assumption including the use of potential fields (${\bf j}= {\bf
0}$) (e.g.\ \opencite{schmidt64}; \opencite{semel67};
\opencite{schatten69}; \opencite{sakurai82};
\opencite{rudenko01a}), linear force-free fields (e.g.\
\opencite{nakagawa72}; \opencite{chiu77}; \opencite{seehafer78};
\opencite{semel88}; \opencite{gary89}; \opencite{lothian95}) and
nonlinear force-free fields (e.g.\ \opencite{sakurai81};
\opencite{wu85}; \opencite{roumeliotis96}; \opencite{amari97};
\opencite{mcclymont97}; \opencite{wheatland00}; \opencite{yan00}).

Ideally, the information contained in a (perfect) vector
magnetogram together with the force-free condition would be
sufficient to calculate the coronal magnetic field. However,
despite the increasing availability of vector magnetogram data, we
are still far from this ideal situation due to both the quality of
the data and to the difficulty of the calculation. It is therefore
still easier to use line-of-sight magnetograms as input for
potential or force-free extrapolation.

Potential fields are completely determined by fixing the
line-of-sight component of the magnetic field \cite{semel67}, but
do not necessarily give good fits to observed emission structures.
In the case of linear force-free fields, the normal (e.g.\
\opencite{chiu77}) or line-of-sight component \cite{semel88} is
not sufficient to determine the field uniquely and one has the
freedom to choose a value for the linear force-free parameter
$\alpha$. Even though some methods have been suggested to
determine $\alpha$ by using the ambiguity of the full linear
force-free solution to fit vector magnetogram data \cite{amari97,
wheatland99}, the usual method is to try to choose the value for
$\alpha$ in such a way that a subset of the field lines matches
the observed emission pattern as good as possible (e.g.\
\opencite{pevtsov95}).

So far, our information about the emitting plasma structures has
been largely limited to two-dimensional projections of
intrinsically three\--dimen\-sio\-nal objects. However, within the
next few years the STEREO mission will hopefully give us the
possibility to get three\--di\-men\-si\-onal information about the
coronal plasma structures, in addition to magnetogram data which
are routinely taken by ground- or space-based instruments.

In the present paper we want explore the possibility to determine
a value of $\alpha$ for a linear force-free field by comparing the
reconstructed magnetic field to the observed
three\--di\-men\-sio\-nal structure of coronal plasma loops. This
is done by defining one or several three-dimensional curves in
space representing the spatial structure of the observed loops,
and a mathematical measure of the deviation of the reconstructed
magnetic field lines from these three\--di\-men\-sio\-nal space
curves. The value of $\alpha$ is then determined by minimising the
deviation from the observed loops. We emphasise that this is to be
considered as a first step only and that a generalisation of the
method to non-linear force-free fields is planned. As the STEREO
mission is yet to be launched, the method is tested using the loop
shapes deducted by \inlinecite{aschwanden99} from applying dynamic
stereoscopy to the solar active region NOAA 7986 using SOHO/MDI
and SOHO/EIT data taken on 29, 30 and 31 August 1996.

The outline of the paper is as follows. In section \ref{program}
we describe the basic algorithm of the reconstruction method. A
brief description of the method used to calculate the linear
force-free fields is given in section \ref{constalpha}. The code
is then applied to the data of \inlinecite{aschwanden99} in
section \ref{applications}. Conclusions and an outlook to future
research is given in section \ref{conclusions}.

\section{The reconstruction method}
\label{program}

Our aim is to develop a method of coronal magnetic field
reconstruction based on linear force-free fields which uses as
input photospheric line-of-sight magnetograms and which optimises
the value of the force-free parameter $\alpha$ in such way that
the resulting field fits the three\--di\-men\-sio\-nal shape of a
coronal loop.

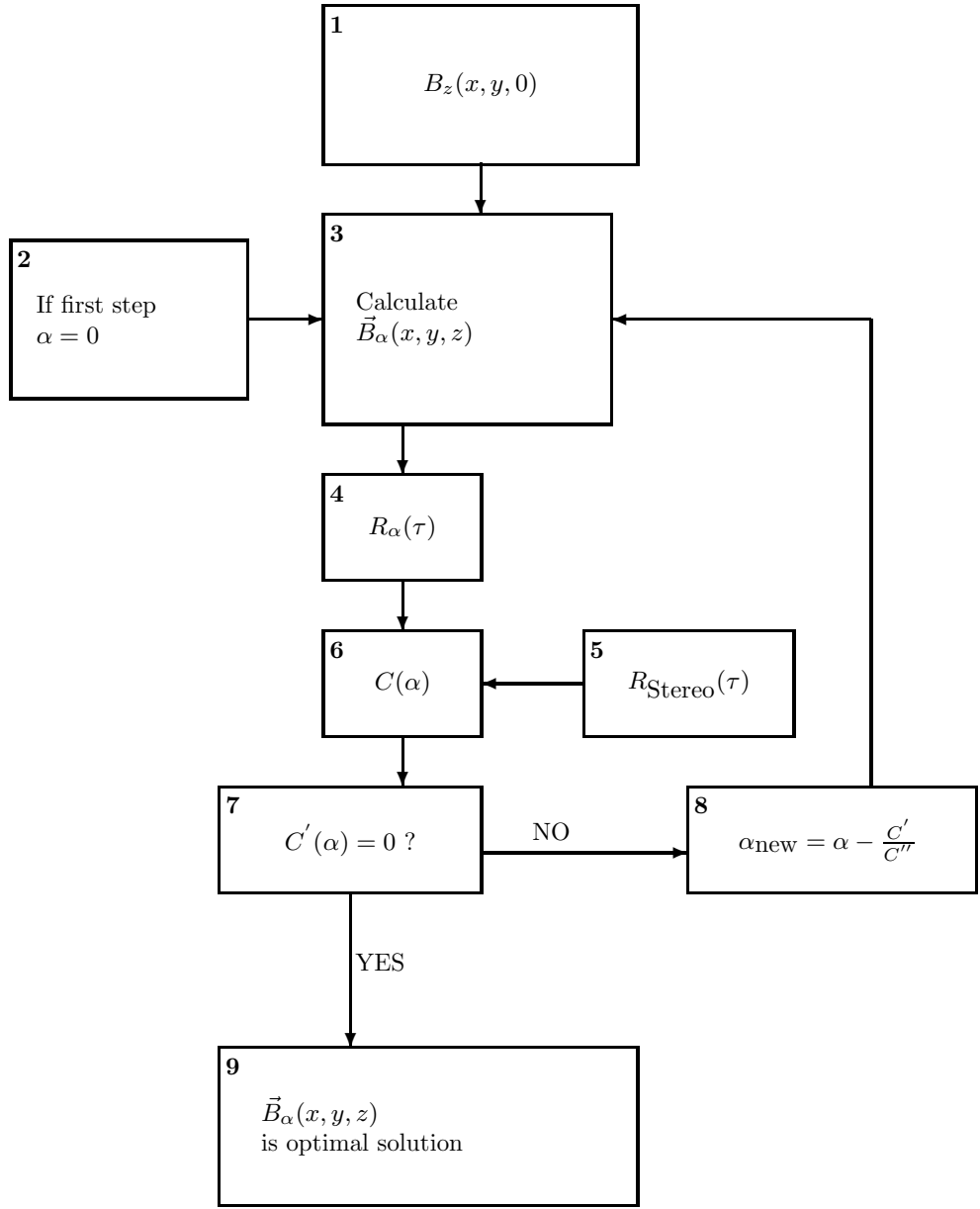
\begin{figure}
\begin{center}
\setlength{\unitlength}{0.7cm}
\begin{picture}(19,24)
\thicklines \put(6,20){\framebox(6,3){$B_z(x,y,0)$}}
\put(6.1,22.5){\bf 1} \put(9,20){\vector(0,-1){1}}
\put(0,15.5){\framebox(4.5,3){\parbox{2.5cm}{{If first step}\\
$\alpha=0$}}} \put(0.1,18.0){\bf 2}

\put(4.5,17){\vector(1,0){1.5}}
\put(6,15){\framebox(5.5,4){\parbox{3.0cm}{Calculate \\ $\vec
B_{\alpha}(x,y,z)$ }}} \put(6.1,18.5){\bf 3}
\put(7.5,15){\vector(0,-1){1.0}}
\put(6,12){\framebox(3,2){$R_{\alpha}(\tau)$}} \put(6.1,13.5){\bf
4}

\put(7.5,12){\vector(0,-1){1.0}}
\put(6,9){\framebox(3,2){$C(\alpha)$}} \put(6.1,10.5){\bf 6}

\put(7.5,9){\vector(0,-1){1.0}}
\put(11,9){\framebox(4,2){$R_{\mbox{{Stereo}}}(\tau)$}}
\put(11.1,10.5){\bf 5}

\put(11,10){\vector(-1,0){2.0}}
\put(4,6){\framebox(5,2){$C^{'}(\alpha)=0$ ?}} \put(4.1,7.5){\bf
7}

\put(6.5,6){\vector(0,-1){3.0}} \put(6.6,4.5){YES}
\put(9,6.75){\vector(1,0){4.0}} \put(10,7){NO}
\put(4,0){\framebox(8,3){\parbox{4.5cm}{$\vec B_{\alpha}(x,y,z)$
\\is optimal solution}}} \put(4.1,2.5){\bf 9}

\put(13,6){\framebox(5.5,2){$\alpha_{\mbox{{new}}}= \alpha-\frac{
C^{'}}{C^{''}}$ }} \put(13.1,7.5){\bf 8}

\put(16.5,8){\line(0,1){9.0}} \put(16.5,17){\vector(-1,0){5.0}}
\end{picture}
\end{center}
\caption{Schematic diagram of algorithm} \label{algorithm}
\end{figure}
The presently planned use of the SECCHI instrument will allow the
reconstruction of loops, or sections of loops, as curves in 3D
space in the following way (B. Inhester, SECCHI Team, private
communication). Once identified in each of the simultaneous SECCHI
images, each image of a loop projected onto the viewing direction
defines a surface on which the sources of the emission must lie.
The intersection of the surfaces from both images then yields once
more 3D curves as the solution to the stereoscopic reconstruction
problem. The solution obviously may not be unique so that
additional information has to be taken into account to select
those curves along which physically meaningful field lines may be
oriented. The benefit of the magnetic field reconstruction method
that we intend to develop is thus two-fold: For once it yields the
full magnetic field structure of those stereoscopic solutions that
are acceptable. Secondly it helps to eliminate unphysical multiple
solutions if a force-free magnetic field cannot be reconstructed
in accordance with the photospheric magnetic field.
The method followed in this paper is outlined in the diagram shown
in figure \ref{algorithm}. The different steps used in this method
are shown as boxes in figure \ref{algorithm}. The details of these
steps used in the method are the following.  We describe the
method only for a single observed loop. The generalisation to
several loops is straightforward and will be briefly explained in
section \ref{multiple}. The numbers of the steps correspond to the
numbers in the boxes in figure \ref{algorithm}. For the iterative
algorithm to work we need a three\--di\-men\-sio\-nal space curve
${\bf R}_{\mbox{Stereo}}(\tau)$ representing the loop shape.

 Here $\tau$ is a parameter which has the value
$\tau=0$ at one foot-point of the loop and the value $\tau=L_l$ at
the other foot-point if the complete loop shape is known. If only
a section of the loop is known the value $\tau=0$ should
correspond to one end point of this section (not necessarily a
foot point) and $\tau=L_l$ to the other end point of the observed
section of the loop. Then $L_l$ does of course not refer to the
total loop length in this case, but only to the length of the
observed section.

We emphasize that the use of $\tau$ as a parameter is only one way
of parametrizing the space curve, in this case using a multiple of
the loop arc length. For comparison of ${\bf
R}_{\mbox{Stereo}}(\tau)$ with other 3D space curves it makes
sense to use the same parameter for the other space curves because
it ensures that any measure of the distance (in the mathematical
sense of a norm) between two curves will vanish if the curves
coincide. Of course, the parametrisation of the other 3D space
curves is in principle completely arbitrary, but we believe that
for the methods described in this paper our choice is the most
practicable one.

\begin{enumerate}
\item The observed photospheric
(line-of-sight) magnetic field $B_z(x,y,0)$ is used as boundary
condition for the magnetic field calculation.
\item
We start by calculating the potential field ($\alpha=0$)
corresponding to the given magnetogram.
\item At later steps during the iteration
$\alpha$ will be non-zero and we have to calculate the linear
force-free field corresponding to the value of $\alpha$ and the
boundary condition given by the (line-of-sight) magnetogram. We
use the method of \inlinecite{seehafer78} for determining the
linear force-free field. Details are described in section
\ref{constalpha}.
\item  Based on the magnetic field ${\bf B}_{\alpha}(x,y,z)$
calculated in step 3, we calculate a field line ${ \bf
R}_{\alpha}(\tau)$ starting on or close to the given stereoscopic
loop. In this paper we have taken either the top of the observed
loop or one of the foot-points as starting point for the
integration. We emphasize that the method will also work if the
known section of the observed loop does not start at the
photosphere. In this case the integration starts at one end point
of the observed loop section and ends at the other end point. The
arc length of the calculated field-line is then normalised to a
fixed value $L_l$ corresponding to the length of the observed
space curve.
\item The information about the observed loop in the corona is provided
in the form of a three\--di\-men\-sio\-nal space curve ${\bf
R}_{\mbox{Stereo}}(\tau)$. The arc length of the loop is
normalised to the same value ($L_l$) as the field-line calculated
in step 4.
\item In this step  the quality of the reconstructed field is assessed by
comparing the observed loop and the reconstructed magnetic field.
In the ideal both space curves would be identical, but this cannot
be generally expected due to possible errors in the observations
and the fact that only a linear force-free field is used here.

In this paper we use two different methods to assess the quality
of the reconstructed field and determine an optimal value for
$\alpha$~:

\begin{enumerate}

\item A simple way to compare the two space curves
is to use one observed foot-point as start-point for calculating
${\bf R}_{\alpha}(\tau)$ and to determine the $\alpha$ which
minimises the distance between the second foot-point of ${\bf
R}_{\alpha}(\tau)$ and ${\bf R}_{\mbox{Stereo}}(\tau)$, i.e. to
minimise $f(\alpha)= \left|{\bf R}_{\alpha}(\tau=L_l)-
                   {\bf R}_{\mbox{Stereo}}(\tau=L_l) \right|$
(Eric Priest, private communication). Whereas this ensures that
one of the foot-points of the two space curves is identical and
the other foot-point as close as possible to the observed
location, the curves as such could have totally different shapes.

This method works only if complete loops including the foot points
are actually known.

\item A more sophisticated way to optimise the
magnetic field is to compare the full 3D structure of the observed
and one or several reconstructed space curves (field lines). The
reconstructed space curves ${\bf R}_{\alpha}(\tau)$ are determined
by first selecting one or several starting points for the field
line integration and then integrating the field lines passing
through these points numerically. The starting points for the
integration can be either points on the observed loop, e.g. the
top or the end points of the observed curve, or other points close
by. We then compare the calculated and the observed curves by
integrating their spatial distance along the complete length of
the curves from $\tau=0$ to $\tau=L_l$ leading to

\[
C(\alpha)=\frac{1}{L_l^2} \, \int\limits_0^{L_l} \sqrt{\left( {\bf
R}_{\mbox{Stereo}}(\tau)-{\bf R}_{\alpha}(\tau)\right)^2} d \tau .
\]
The value of $\alpha$ for the best fitting field line is then
determined by minimising $C(\alpha)$. If several field lines are
compared with the same observed curve, the total minimum of all
field lines is chosen. A possible variant not used in the present
paper is to minimise the sum of all individual $C(\alpha)$ for a
set of field lines.

This method can also be applied to loop sections in cases where
the complete loop has not been observed. It should be noted,
however, that the method becomes less and less meaningful with
decreasing length of the observed loop sections.

$C(\alpha)$ as defined above is dimensionless and its value also
provides a measure of how much the reconstructed field line and
the observed space curve differ for a given value of $\alpha$. We
have normalised $C(\alpha)$ to by the length of the observed loop
(or loop section) so that the values of $C$ should be more or less
independent of the loop length. In this case values of $C$ of the
order of or less than unity indicate good fits, whereas higher
values of $C(\alpha)$ indicate bad fits.

It can be useful to calculate $C(\alpha)$ for different values of
$\alpha$ to determine suitable initial values for a Newton
iteration to determine the minimum of $C(\alpha)$ (see next step).
\end{enumerate}
\item The optimal value for $\alpha$ is determined by minimising $C(\alpha)$.
This is done by calculating the zeros of $C^{'}(\alpha)$. As we
are interested in the absolute minimum of $C(\alpha)$ the
knowledge gained in the previous step is very helpful to see
whether there are several minima, and which values of $\alpha$ are
useful as starting points.

$C^{\prime}(\alpha)$ and $C^{\prime\prime}(\alpha)$ are calculated
numerically. If the current $\alpha$ minimises $C(\alpha)$ (YES
arrow in figure \ref{algorithm}) the optimal linear force-free
solution has been found.

If the minimum has not been found to within the desired accuracy
(NO arrow in figure \ref{algorithm}), the next iteration step is
carried out (see step 8).

\item
A new value for $\alpha$ is determined by a Newton-Raphson
iteration step
\[
\alpha_{n+1}=
\alpha_n-\frac{C^{\prime}(\alpha_n)}{C^{\prime\prime}(\alpha_n)}.
\]
This new value for $\alpha$ is then used as input for the field
solver (step 3).

\end{enumerate}
The iteration is continued until $C(\alpha)$ has been minimised to
the desired degree of accuracy.

The resulting magnetic field can be considered as the optimal
linear force-free magnetic field under the constraints that it
satisfies the boundary conditions given by the magnetogram and
that it minimises the difference between a particular field line
and the observed loop shape.

\section{The linear force-free field solver}

\label{constalpha}

We use the method of \inlinecite{seehafer78} for calculating the
linear force-free field for a given magnetogram and a given value
of $\alpha$. This method gives the components of the magnetic
field for a semi-finite column of rectangular cross-section in
terms of a Fourier series.

The observed magnetogram which covers a rectangular region
extending from $0$ to $L_x$ in $x$ and $0$ to $L_y$ in $y$ is
artificially extended onto a rectangular region covering $-L_x$ to
$L_x$ and $-L_y$ to $L_y$ by taking an antisymmetric mirror image
of the original magnetogram in the extended region, i.e.
\begin{eqnarray*}
B_z(-x,y) &= &-B_z(x,y)\\ B_z(x,-y) &= &-B_z(x,y).
\end{eqnarray*}
The advantage of taking the antisymmetric extension of the
original magnetogram is that the extended magnetogram is
automatically flux balanced. The method has the further advantage
that a Fast Fourier Transformation (FFT) scheme (see also
\opencite{alissandrakis81}) can be used to determine the
coefficients of the Fourier series. For more details regarding
this method see \inlinecite{seehafer78}, and a comparison of the
performance of the method with other reconstruction methods has
been given by \inlinecite{seehafer82}.

The expression for the magnetic field is given by
\begin{eqnarray}
B_x & = & \sum_{m,n=1}^{\infty} \frac{C_{mn}}{\lambda_{mn}} \exp
\left(-r_{mn} z \right) \cdot
 \left[ \alpha \frac{\pi n}{L_y} \sin \left(\frac{\pi m x}{L_x}\right)
\cos \left(\frac{\pi n y}{L_y}\right) - \right. \nonumber \\
&&\left. -r_{mn} \frac{\pi m}{L_x} \cos \left(\frac{\pi m
x}{L_x}\right) \sin \left(\frac{\pi n y}{L_y}\right) \right] \\
B_y & = & -\sum_{m,n=1}^{\infty} \frac{C_{mn}}{\lambda_{mn}} \exp
\left(-r_{mn} z \right) \cdot
 \left[ \alpha \frac{\pi m}{L_x} \cos \left(\frac{\pi m x}{L_x}\right)
\sin \left(\frac{\pi n y}{L_y}\right)                 +
\right. \nonumber \\ && \left.+r_{mn} \frac{\pi n}{L_y} \sin
\left(\frac{\pi m x}{L_x}\right) \cos \left(\frac{\pi n
y}{L_y}\right) \right] \\
B_z & = & \sum_{m,n=1}^{\infty} C_{mn} \exp \left(-r_{mn} z
\right) \cdot \sin \left(\frac{\pi m x}{L_x}\right) \sin
\left(\frac{\pi n y}{L_y}\right)
 \label{B_z}
\end{eqnarray}
with $\lambda_{mn}= \pi^2 (m^2/L_x^2+ n^2/L_y^2 )$ and
$r_{mn}=\sqrt{\lambda_{mn}-\alpha^2}$.

The coefficients $C_{mn}$ are obtained by comparing Equation
(\ref{B_z}) for $z=0$ with a FFT of the magnetogram data. The
numerical method has to cut-off the Fourier series at some maximum
values for $m_{\mbox{max}}$ and $n_{\mbox{max}}$. For the example
present in section \ref{applications}
$m_{\mbox{max}}=n_{\mbox{max}}=40$ was used. Further Fourier
coefficients could be taken into account if very small scale
structures have to be resolved, which is not the case in the
present paper.

We use the SI-system throughout the paper, with the exception of
the magnetic field strength to which we refer in Gauss ($1$ Gauss
= $10^{-4}$ Tesla). Due to the antisymmetry of the extended
magnetogram the first term contributing to the magnetic field is
the $m=n=1$ term. Therefore the maximum value of $\alpha^2$ for
given $L_x$ and $L_y$ is
\[
\alpha^2_{\mbox{max}}= \pi^2\left(\frac{1}{L_x^2}
+\frac{1}{L_y^2}\right).
\]
To normalise $\alpha$ we choose the harmonic mean $L$ of $L_x$ and
$L_y$ defined by
\[
\frac{1}{L^2}=\frac{1}{2}\left(
\frac{1}{L_x^2}+\frac{1}{L_y^2}\right) .
\]
For $L_x=L_y$ we have $L=L_x=L_y$. With this normalisation the
values of $\alpha$ fall into the range $-\sqrt{2} \pi < \alpha <
\sqrt{2} \pi$.

We would like to emphasize that our optimisation method does not
rely on the way the linear force-free field is calculated, i.e.
any other method, for example a Green's function method, can be
used as well. We have chosen the Seehafer (1978) method only for
computational convenience.

\section{Applications}
\label{applications}

\subsection{Application to a single loop}

\label{single}

In principle the method described in section \ref {program} could
be tested by using any reasonable three\--di\-men\-sio\-nal space
curve as a model loop shape. However, we considered it more
challenging to apply the method to a more realistic situation.
Before the launch of the STEREO mission, true stereoscopic data
will not be available, and we therefore used the loop shapes
determined by \inlinecite{aschwanden99} using the method of
dynamic stereoscopy. Dynamic stereoscopy uses the solar rotation
to get different viewing angles at different observation  times to
derive the three\--di\-men\-sio\-nal loop shapes. The fundamental
assumption of dynamic stereoscopy is that the shapes of the loop
structures vary only very slowly over the period of time of the
observations.

\begin{figure}
\centerline{
\includegraphics[width=8.0cm,height=7.134cm]{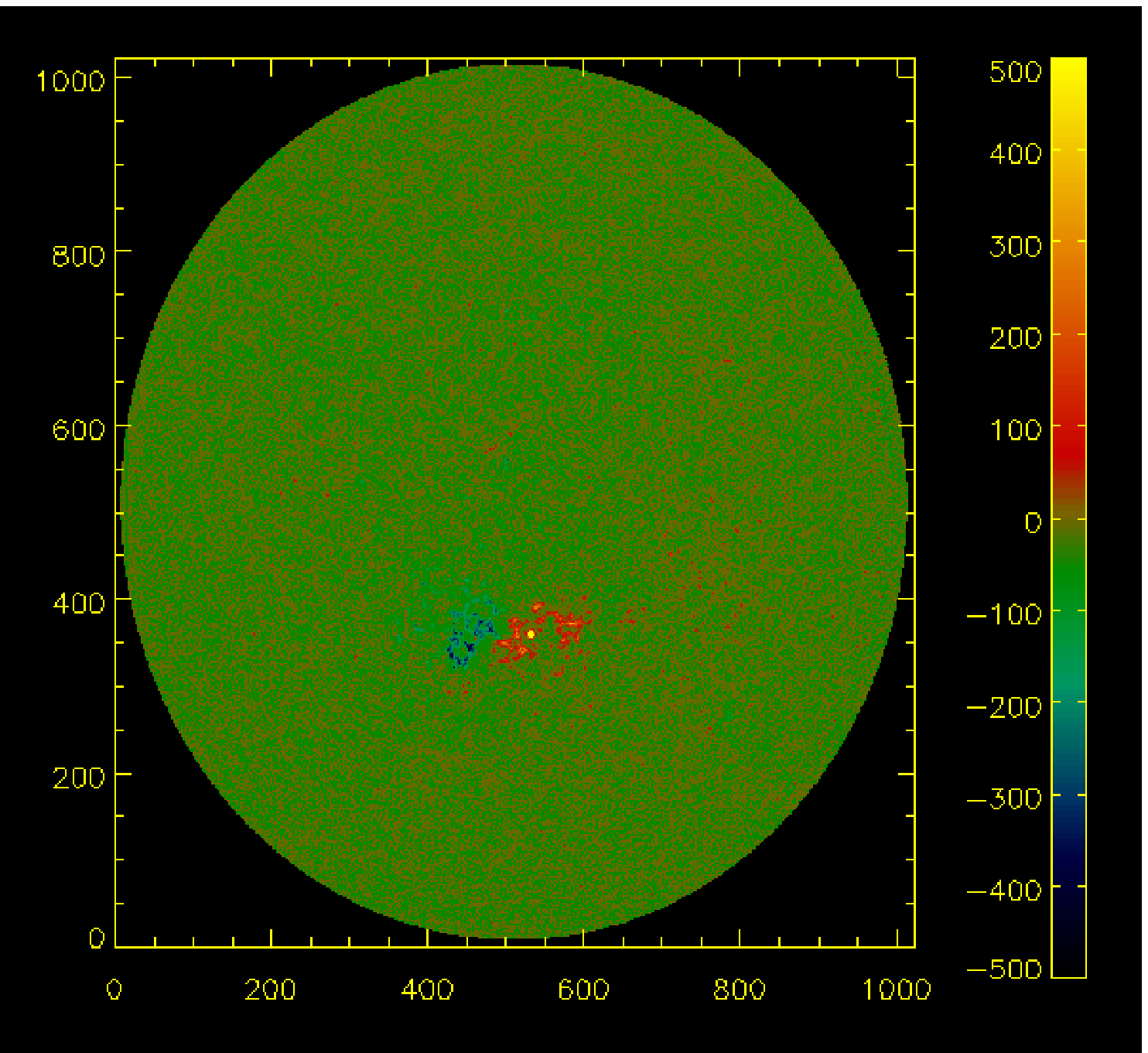}}
\centerline{
\includegraphics[width=8.0cm, height=8.0cm]{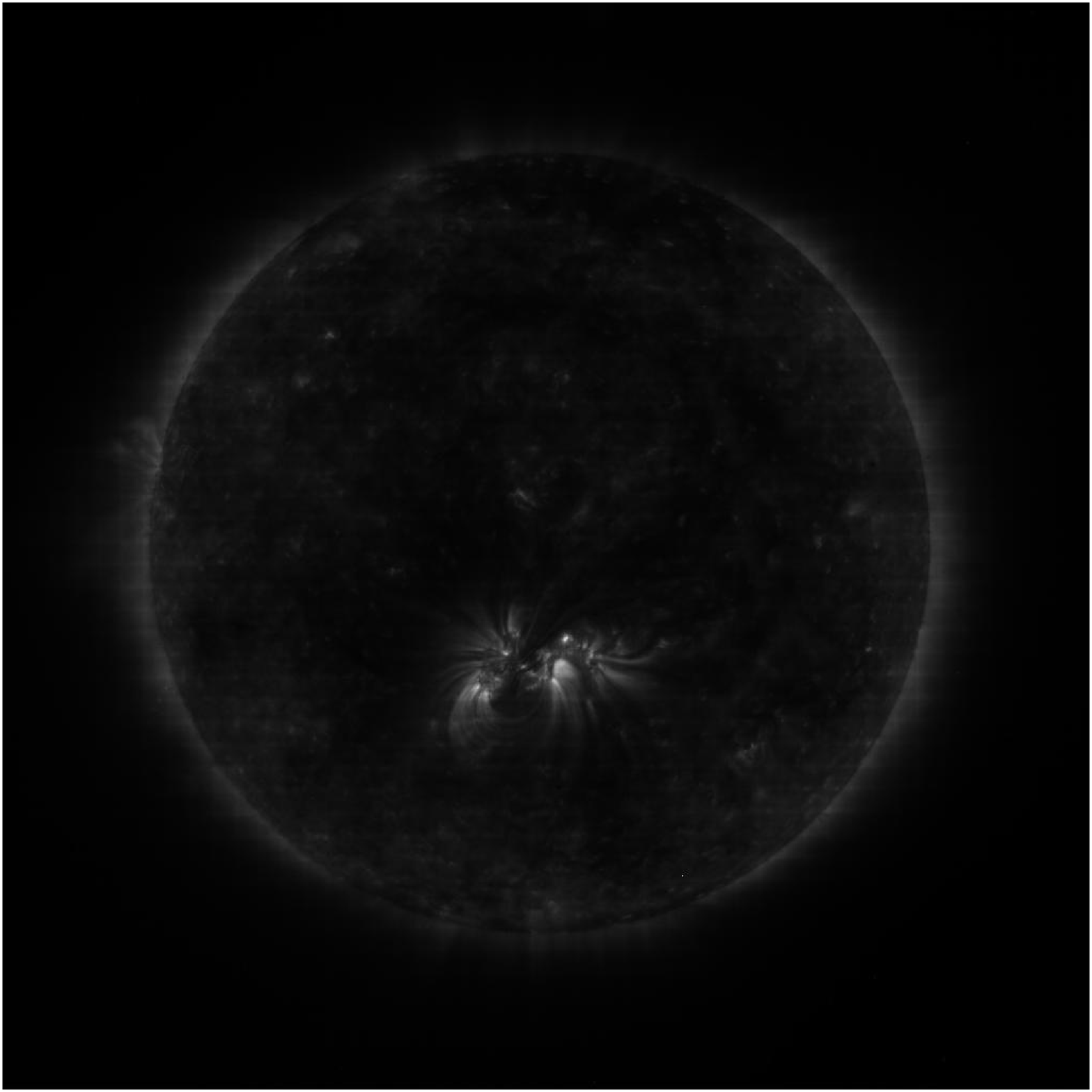}}
\caption{Top: Full disk MDI magnetogram for 30 August 1996,
bottom: Full disk EIT (wavelength171 A) image for 30 August 1996.
The active region NOAA 7986 is clearly visible somewhat below the
disk centre.} \label{mdi-eit}
\end{figure}


\begin{figure}
\centerline{
\includegraphics[width=8.0cm]{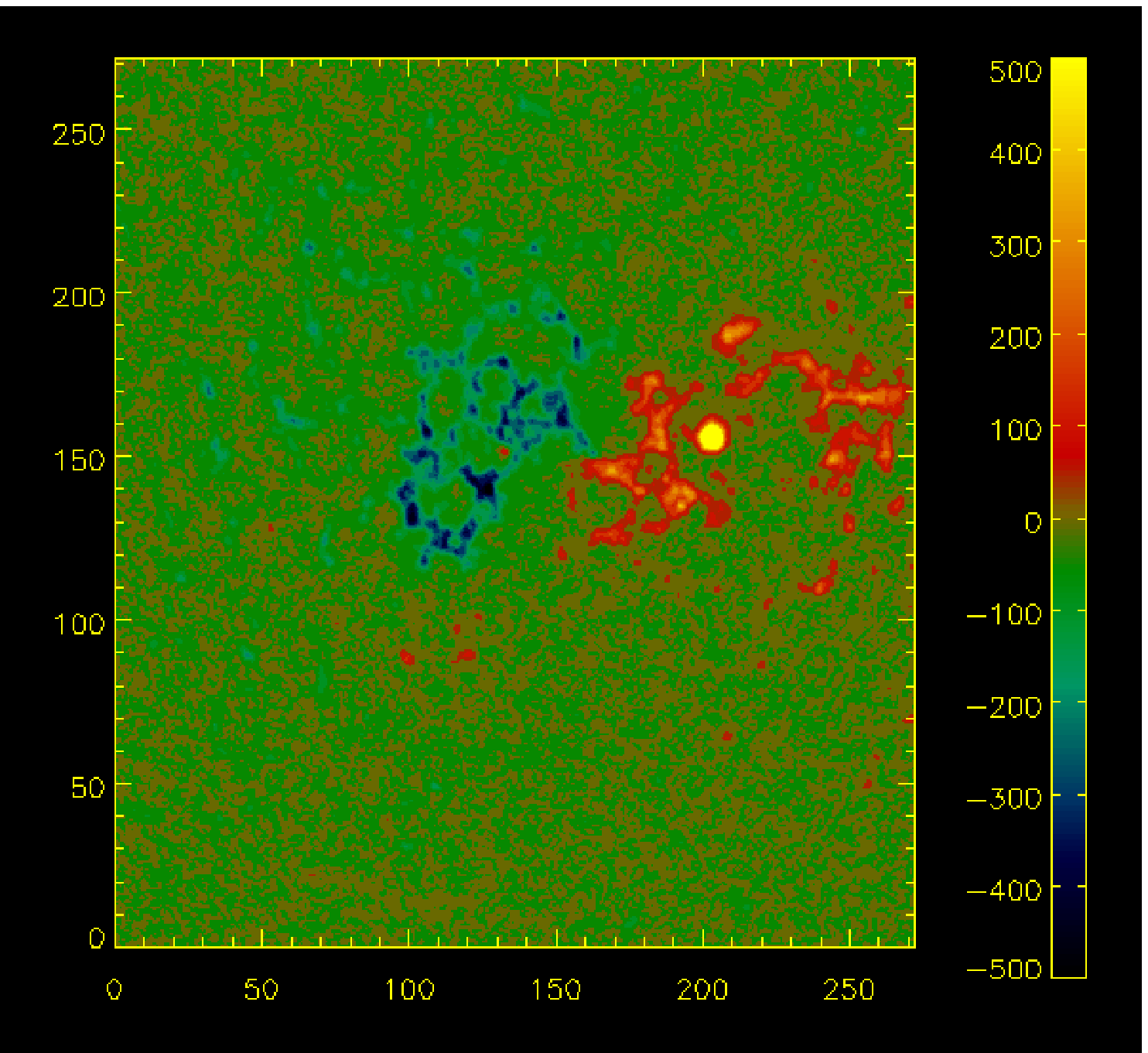}}
\caption{Part of the full disk MDI magnetogram for 30 August 1996
with the active region extracted.} \label{mdipart}
\end{figure}


\inlinecite{aschwanden99} applied the method of dynamic
stereoscopy to the active region NOAA 7986 observed on 30 August
1996 with the EIT and MDI instruments aboard the SOHO spacecraft
(see figure \ref{mdi-eit}). To derive the
three\--di\-men\-sio\-nal loop structure on the 30 August 1996,
\inlinecite{aschwanden99} use EIT observations of the same active
region taken on 29 August 1996 and 31 August 1996.

In total \inlinecite{aschwanden99} reconstructed thirty loops, but
only one loop (Loop 1 in \inlinecite{aschwanden99}) was traced
along its whole length. It seems natural to choose this loop as a
reference case to check the capabilities of the reconstruction
method described in section \ref{program} (e.g. iteration of
$\alpha$).

For the reconstruction method we consider only a part of the full
disc magnetogram close to the active region. We use Cartesian
coordinates, with the $z$ axis pointing in the direction
perpendicular to the photosphere. The considered area is shown in
figure \ref{mdipart} and corresponds to the pixels $330 < x < 602
,\quad 203 < y < 475 $ of the full disk $1024 \times 1024$ pixel
MDI-image. The same region has been used in the corresponding
EIT-image by \inlinecite{aschwanden99} for the dynamic
stereoscopy. This corresponds to a normalising length scale for
$\alpha$ of $L=L_x=L_y= 385$ Mm. The corresponding values of
$\alpha$ in SI units are $|\alpha| < 1.154 \cdot 10^{-8}$
m$^{-1}$.

We applied both the method of minimising the distance of
foot-points and the preferable method of minimising the distance
of loop segments between the observed and calculated loop. In the
first case we start the field line integration at one of the foot
points of the observed loop, in the second case  we start the
field line integration at the top of the observed loop and
integrate in both directions down to the photosphere ($z=0$).

\begin{figure}
\centerline{
\includegraphics[width=6.0cm]{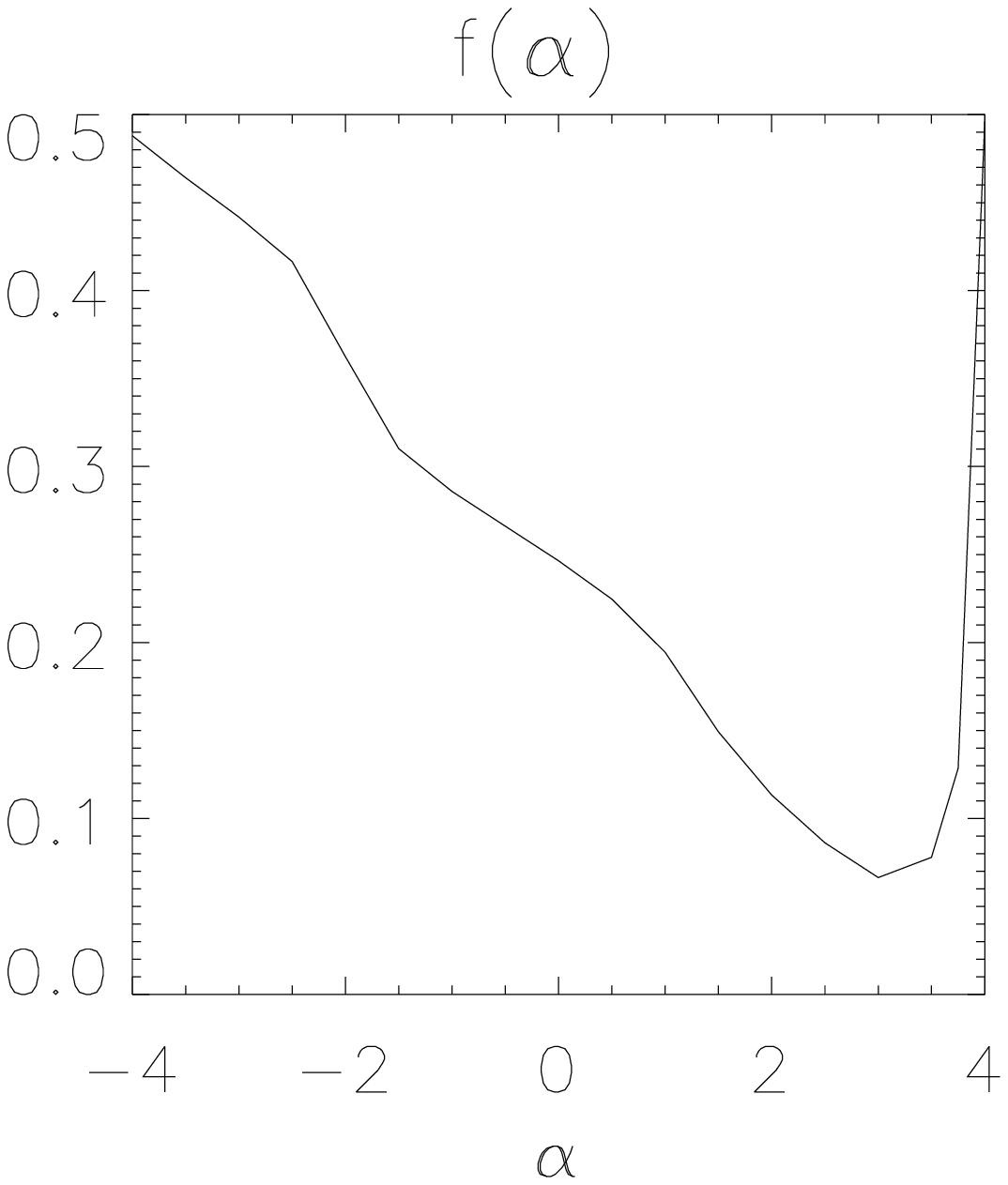}
\includegraphics[width=6.0cm]{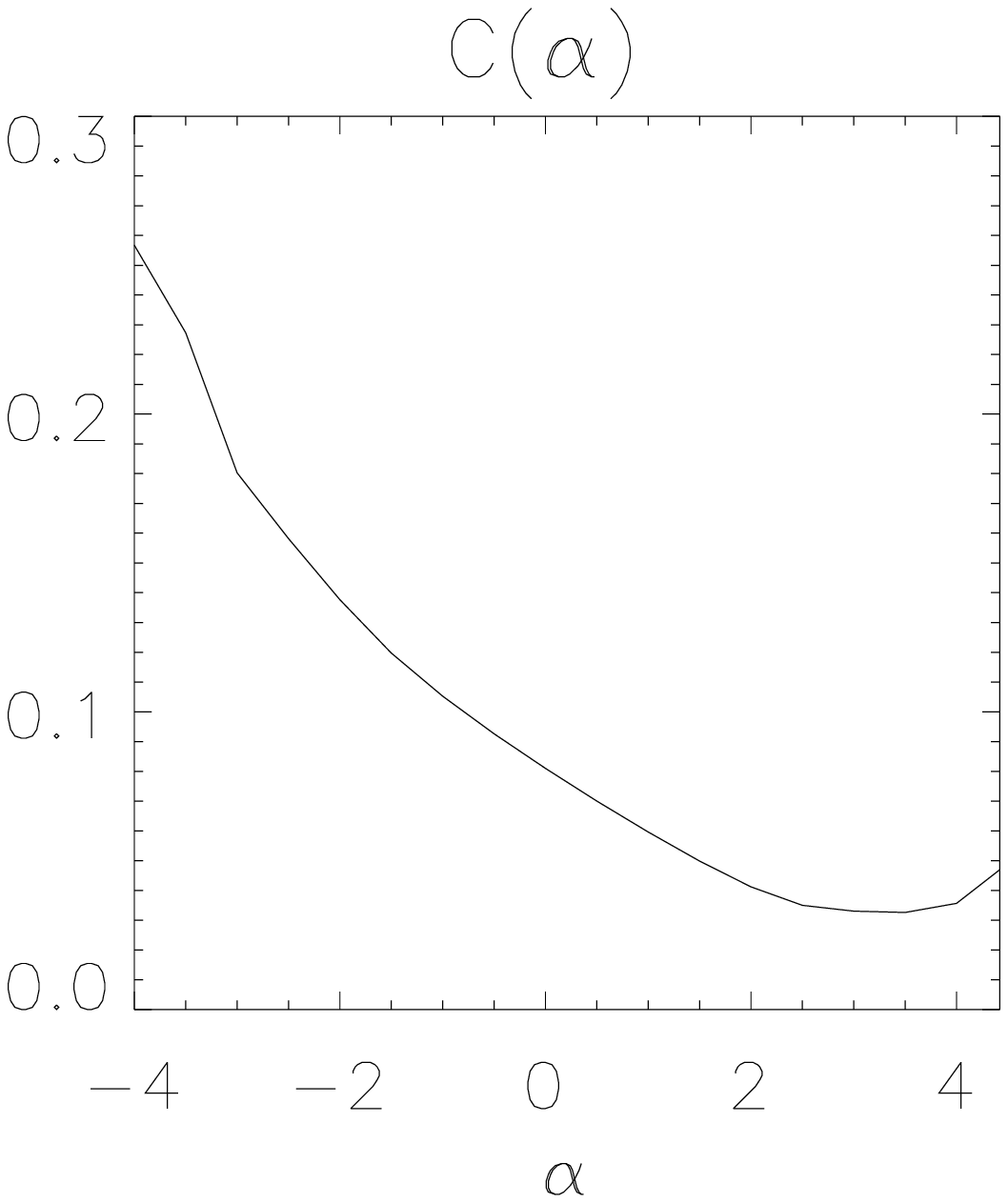}
} \caption{The functions $f(\alpha)$ (left) and $C(\alpha)$
(right) for loop 1 of \protect\inlinecite{aschwanden99}. The
function $f(\alpha)$ (foot point distance method) has a minimum at
$\alpha=3.0$ and $C(\alpha)$ (loop distance method) has a minimum
at $\alpha=3.5$. The rapid increase of the functions, in
particular of $f(\alpha)$, for larger $\alpha$ values is caused by
the rapid change of the magnetic field as the limiting value of
$\alpha=\sqrt{2}\pi\simeq 4.44$ is approached.} \label{cvona}
\end{figure}

Figure \ref{cvona} shows the corresponding functions $f(\alpha)$
for the foot-point method on the left-hand side and and $
C(\alpha)$ for the loop distance method on the right-hand side.
The minimum of $f(\alpha)$ occurs at $\alpha=3.0$ and the minimum
of $C(\alpha)$ at $\alpha=3.5$. We point out that the minimum for
the $C(\alpha)$ is very flat (see right-hand side in figure
\ref{cvona}) which means that we can expect that values of
$\alpha$ which differ slightly from the minimum value will still
give fits of similar quality. As the limiting value
$\alpha=\sqrt{2}\pi\simeq 4.44$ is approached, both functions show
a rapid increase, in particular of $f(\alpha)$. This is caused by
the rapid change of the magnetic field close the limiting value of
$\alpha$.

\begin{figure}
\centerline{
\includegraphics[width=8.0cm]{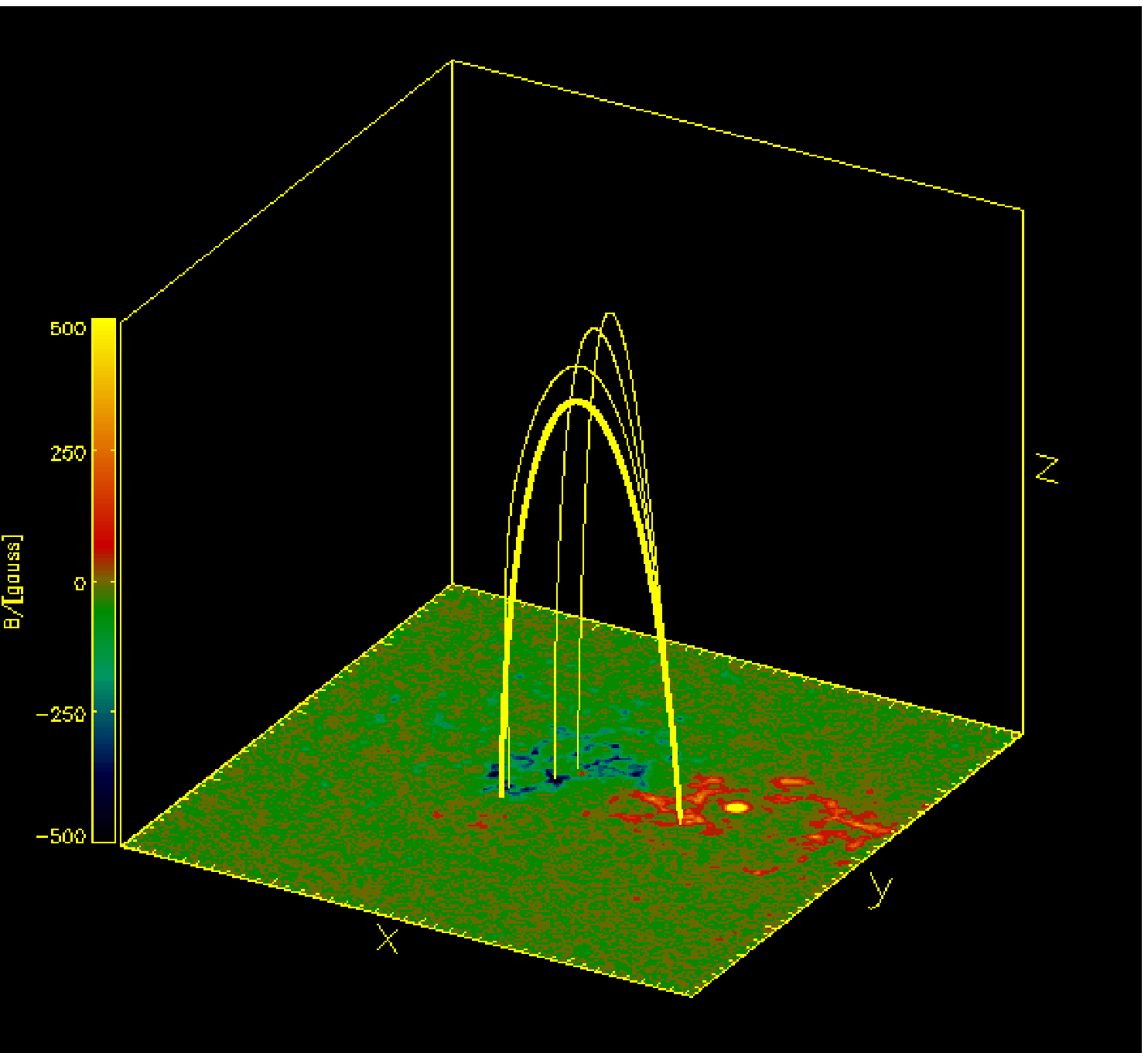}}
\centerline{
\includegraphics[width=8.0cm]{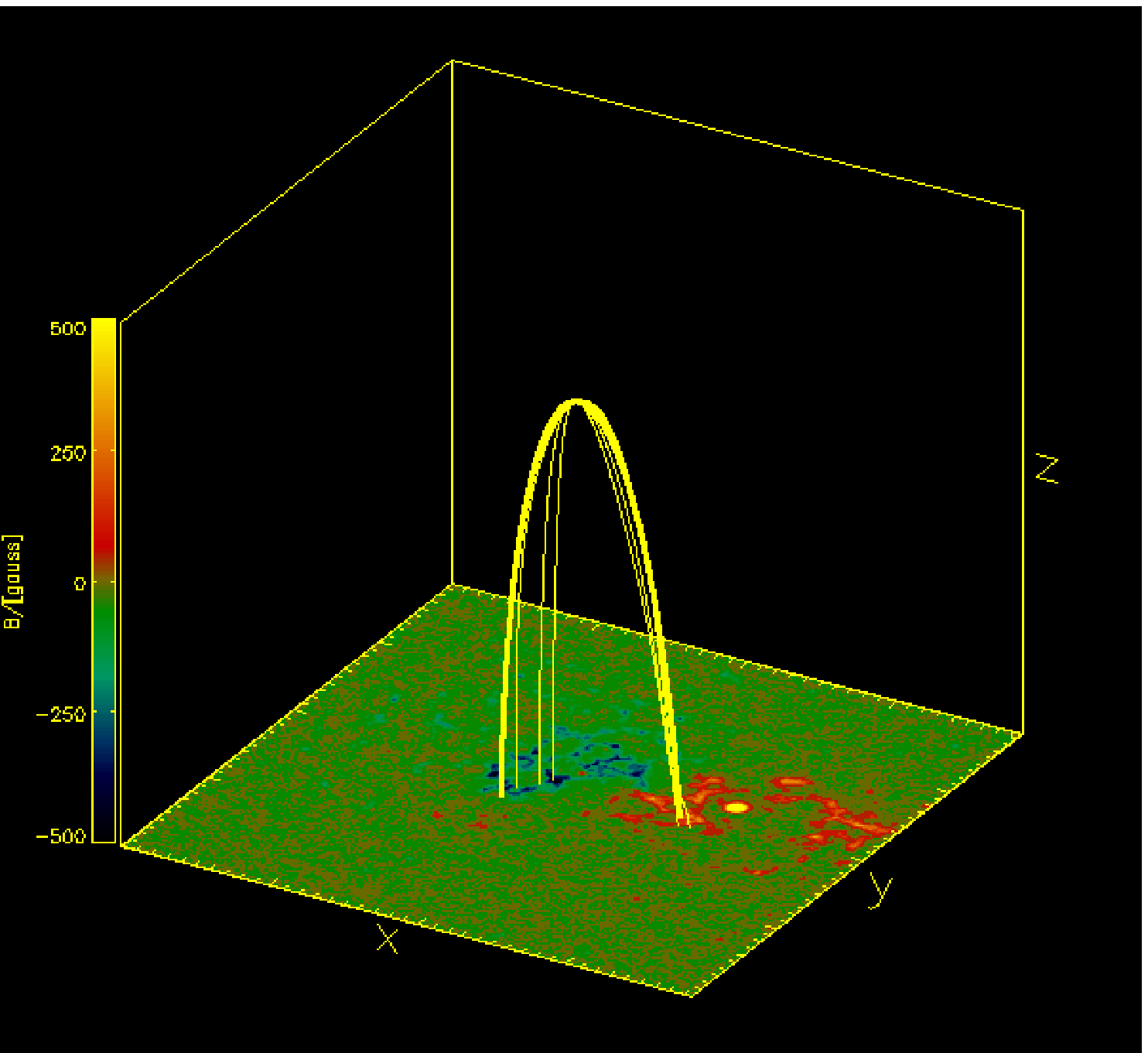}
} \caption{Some field lines for different values of $\alpha$ and
the observed loop (thick line). The $z$-axis has been stretched by
a factor of $5$ to make the comparison of the observed loop and
the field lines easier. In the top panel we have used the foot
point distance method to determine the optimum value $\alpha=3.0$.
The two other field lines shown start at the same foot point, but
are for $\alpha=0.0$ and $\alpha = -2.0$. In the bottom panel the
optimal value $\alpha=3.5$ has been determined by the loop
distance method with starting point of integration at the loop
top. The other field lines shown start at the same point but are
for $\alpha=0.0$ and $\alpha = -2.0$.} \label{fieldlines}
\end{figure}

Figure \ref{fieldlines} shows the loop shape as deduced from the
data (thick line) and field lines of the reconstructed coronal
magnetic field for different values of $\alpha$. The top panel
corresponds to the foot-point method and the bottom panel to the
loop distance method. The reconstructed field-lines for the
optimal values of $\alpha$ nearly coincide with the observed
loops. For the foot-point method the optimal reconstructed loop is
slightly higher than the observed loop. In figure
\ref{fieldlines}, the $z$-axis has been stretched by a factor of
$5$ to facilitate the comparison between the observed loop and the
calculated field lines.

For the loop distance method one observes some deviation between
observed loop and the optimal reconstructed loop at the
foot-points. These deviations are slightly bigger at the left
foot-point. This mismatch of the foot-points has to be balanced
against the fact that the position of the observed end points of
the loop is only accurate to about $10 \%$ of the total observed
length of the loop \cite{aschwanden99}. One also has to account
for the fact that it is not clear whether the observed end points
of the loop coincide with its foot points, i.e. those points where
the field lines of the loop meet the photosphere.

In both cases, we also show field lines with the same starting
points, but values of $\alpha$ which are different from the
optimal value. It is obvious that the match between those field
lines and the observed loop is not as good as that of the optimal
reconstructed field.

In order to see how robust the loop distance method is with
respect to changes in the integration start point of the field
line used for calculating $C(\alpha)$, we have calculated
$C(\alpha)$ for a grid of start points on and around the observed
loop $1$ of \inlinecite{aschwanden99}. In particular, we have
chosen starting points on the observed loop at $1/5$, $2/5$,
$3/5$, and $4/5$ of the observed loop length plus eight further
starting points at each of these distances along the loop, but at
a given distance form the loop position. Four of these eight start
points are located at a distance of $1/50$ of the loop length in a
plane perpendicular to the local tangent of the loop. The angles
between the lines connecting to neighbouring start points with the
loop is $\pi/2$. The other four start points at each of the
distances along the loop are arranged in the same way, but at
twice the distance from the loop. Thus we have $36$ field lines
and we now seek the optimum value of $\alpha$ over the complete
set of field lines.

\begin{figure}
\centerline{
\includegraphics[width=8.0cm]{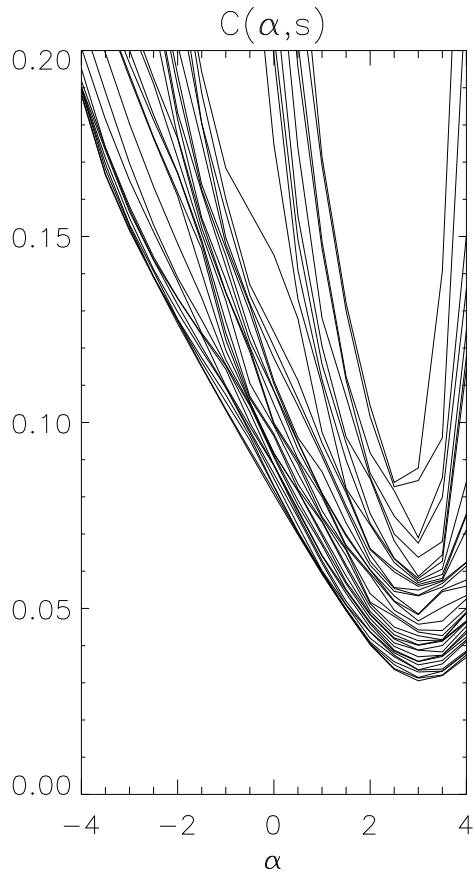}}
\caption{The functions $C_i(\alpha)$ calculated for the $36$
different starting points for field line integration close to the
observed loop $1$ of \protect\inlinecite{aschwanden99}. The
optimum value of $\alpha$ is around $3.0$ in this case. The lower
values of $C_i(\alpha)$ correspond to field lines starting very
close to the top of the observed loop. } \label{cmulti}
\end{figure}
The functions $C_i(\alpha)$, $i=1,\ldots,36$, are shown in figure
\ref{cmulti}. The optimum value of $\alpha$ derived by this method
is approximately $\alpha \simeq 3.0$. The best fitting field lines
all have starting points which are on or very close to the top of
the observed loop. This method gives a value of $\alpha$ which is
very similar to the values previously derived (within a range of
$\pm 0.5$), which indicates that the exact value of the starting
point for the field line integration does not influence the value
of $\alpha$ very much. In particular, figure \ref{cmulti} shows
that the minimum of all $C_i(\alpha)$ is relatively close to
$\alpha \simeq3.0$.

\subsection{Application to multiple loops}

\label{multiple}

We have so far applied the method only to one single loop. If this
particular loop by chance does not represent the generic
properties of the magnetic field, this may lead to a misleading
value for $\alpha$ and a misrepresentation of the field. We
therefore want to extend the methods described in section 2 to
multiple loops. This is easily done by using the sum over several
loops of either $f(\alpha)$ or $C(\alpha)$ in the minimisation
process.

\begin{table}
\caption{Values of $\alpha$  for all $30$ loops (loop number in
first column) by \protect\inlinecite{aschwanden99} determined by
a) applying loop distance method to each loop individually (second
column, value of $C(\alpha)$ in third column), b) applying the
loop distance method to groups of loops (fourth column, value of
$C(\alpha)$ in fifth column), and c) applying the foot point
distance method to individual loops (sixth column, value of
$f(\alpha)$ seventh column). } \label{table1}
\begin{tabular}{|r|r|r|r|r|r|r|}
\hline
       Loop &Optimal $\alpha $&$  C(\alpha)$ &group $\alpha$ & $C(\mbox{group} \
\alpha)$ &
       Optimal $\alpha$&$f(\alpha)$\\
\hline
       1  &$    3.5  $&$    0.033 $ & $2.5$ & $ 0.035$&$3.0$&$  0.066$\\
       2  &$    2.0  $&$    0.026 $ & $2.5$ & $ 0.027$&$2.0$&$  0.048$\\
       3  &$    3.0  $&$    0.058 $ & $2.5$ & $ 0.058$&$2.0$&$  0.043$\\
       4  &$    2.0  $&$    0.076 $ & $2.5$ & $ 0.076$&$2.0$&$  0.017$\\
       5  &$    3.0  $&$    0.030 $ & $2.5$ & $ 0.032$&$2.0$&$  0.019$\\
       6  &$    3.0  $&$    0.027 $ & $2.5$ & $ 0.030$&$2.5$&$  0.015$\\
       7  &$    2.5  $&$    0.081 $ & $2.5$ & $ 0.081$&$2.5$&$  0.027$\\
       8  &$    1.5  $&$    0.312 $ &&&&\\
       9  &$    0.5  $&$    0.077 $ & $0.0$ & $ 0.079$&$1.0$&$ 0.051$\\
       10  &$   2.0  $&$    0.280 $ &&&&\\
       11  &$   3.0  $&$    0.226 $ &&&&\\
       12  &$  -0.5  $&$    0.300 $ &&&&\\
       13  &$  -1.0  $&$    0.048 $ & $-2.0$ & $0.056$&$-1.0$&$  0.010$\\
       14  &$  -0.5  $&$    0.308 $ &&&& \\
       15  &$  -1.5  $&$    0.261 $ &&&& \\
       16  &$  -2.0  $&$    0.077 $ & $-2.0$ & $ 0.077$&$-2.0$&$ 0.059$\\
       17  &$  -2.5  $&$    0.064 $ & $-2.0$ & $ 0.070$&$-2.5$&$ 0.030$\\
       18  &$  -4.0  $&$    0.122 $ &&&&\\
       19  &$  -1.0  $&$    0.228 $ &&&&\\
       20  &$  -3.0  $&$    0.045 $ & $-2.0$ & $ 0.051$&$-3.0$&$ 0.037$ \\
       21  &$  -1.0  $&$    0.043 $ & $-2.0$ & $ 0.045$&$-1.5$&$ 0.038$ \\
       22  &$  -1.0  $&$    0.997 $ &&&&\\
       23  &$   2.5  $&$    0.421 $ &&&&\\
       24  &$  -2.0  $&$    0.411 $ &&&&\\
       25  &$  -4.0  $&$    0.235 $ &&&&\\
       26  &$   0.5  $&$    0.311 $ &&&&\\
       27  &$   2.0  $&$    0.299 $ &&&&\\
       28  &$   1.0  $&$    0.308 $ &&&&\\
       29  &$   1.0  $&$    0.323 $ &&&&\\
       30  &$   1.0  $&$    0.315 $ &&&&\\
       \hline

\end{tabular}
\end{table}

For comparison we first apply the method described in section
\ref{single} to all thirty loops found by
\inlinecite{aschwanden99} individually. A list of the $\alpha$
value for each loop can be found in Table \ref{table1} together
with the value $C(\alpha)$. Due to the very shallow minimum in
$C(\alpha)$ and due to the uncertainties in the determination of
the observed loop shapes, we only determined the $\alpha$ values
in half integer steps.

\inlinecite{aschwanden99} only traced loop $1$ fully and assumed a
circular loop shape for the non-traced parts of the other loops.
For some loops less than $10 \%$ of their length has been traced
(see Table 1 in \inlinecite{aschwanden99}). Certainly force free
loops will usually not be circular, and if the ad hoc assumption
of circular loop shapes is wrong for a poorly traced loop, one
cannot expect a satisfactory a agreement with the results of our
method. However, since in the present paper we are mainly
interested in testing our method we have therefore included the
results for poorly traced loops in Table \ref{table1}. We
considered values of $C < 0.1 $ (in the normalisation used) as
small enough, because checking the fits to the observed loops
still gave acceptable results for this value. The values of
$\alpha$ for all other loops have to be regarded with caution.

In this sense, we are able to fit $13$ of the $30$ loops with
reasonable accuracy, as indicated by relatively small values of
the individual $C(\alpha)$. We emphasize that our assumption of
linear force-free fields may also be too restrictive for some
loops. This once more indicates the necessity to extend the method
to nonlinear force free fields.

\begin{figure} 
\centerline{
\includegraphics[width=8.0cm]{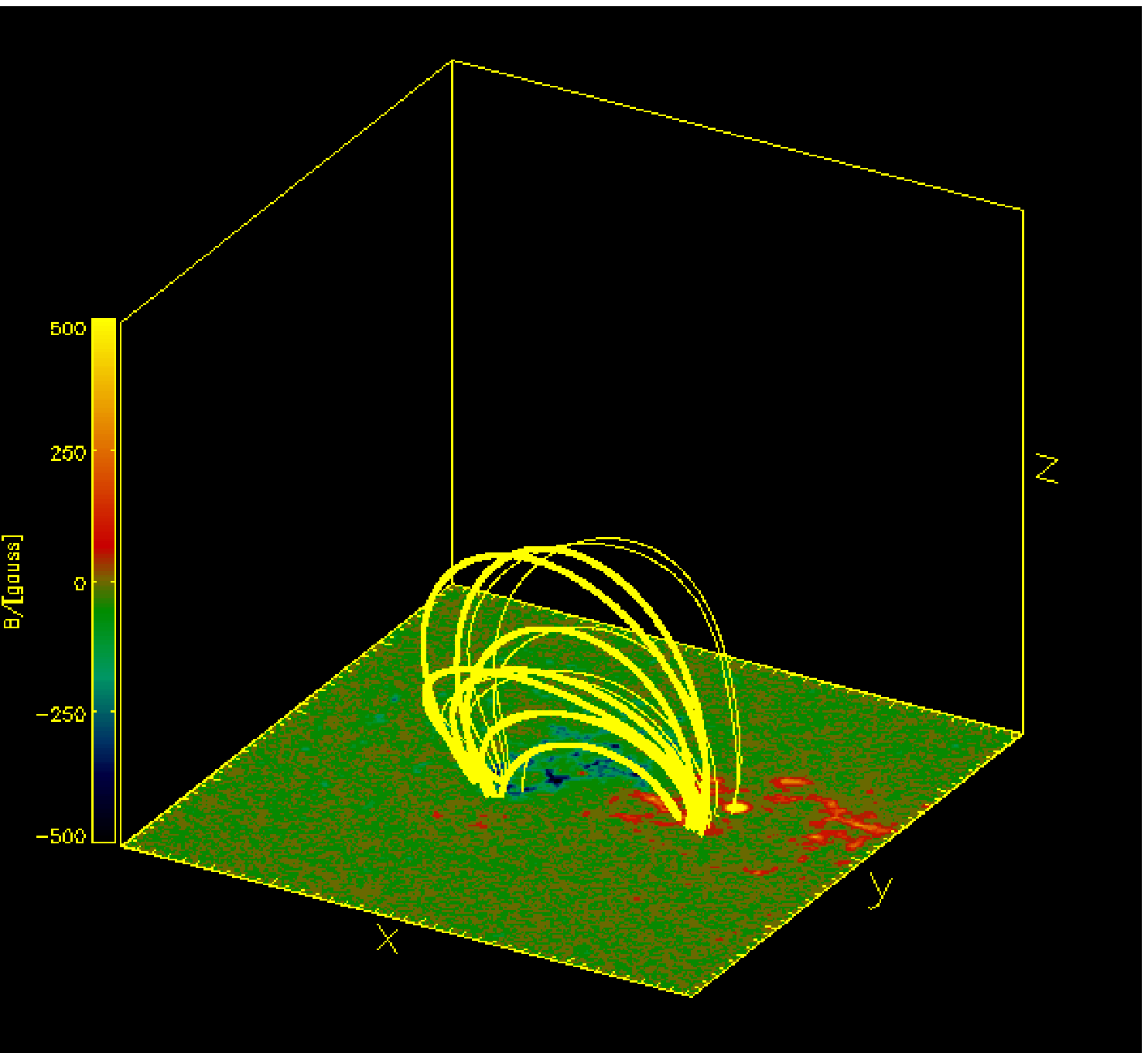}}
\centerline{
\includegraphics[width=8.0cm]{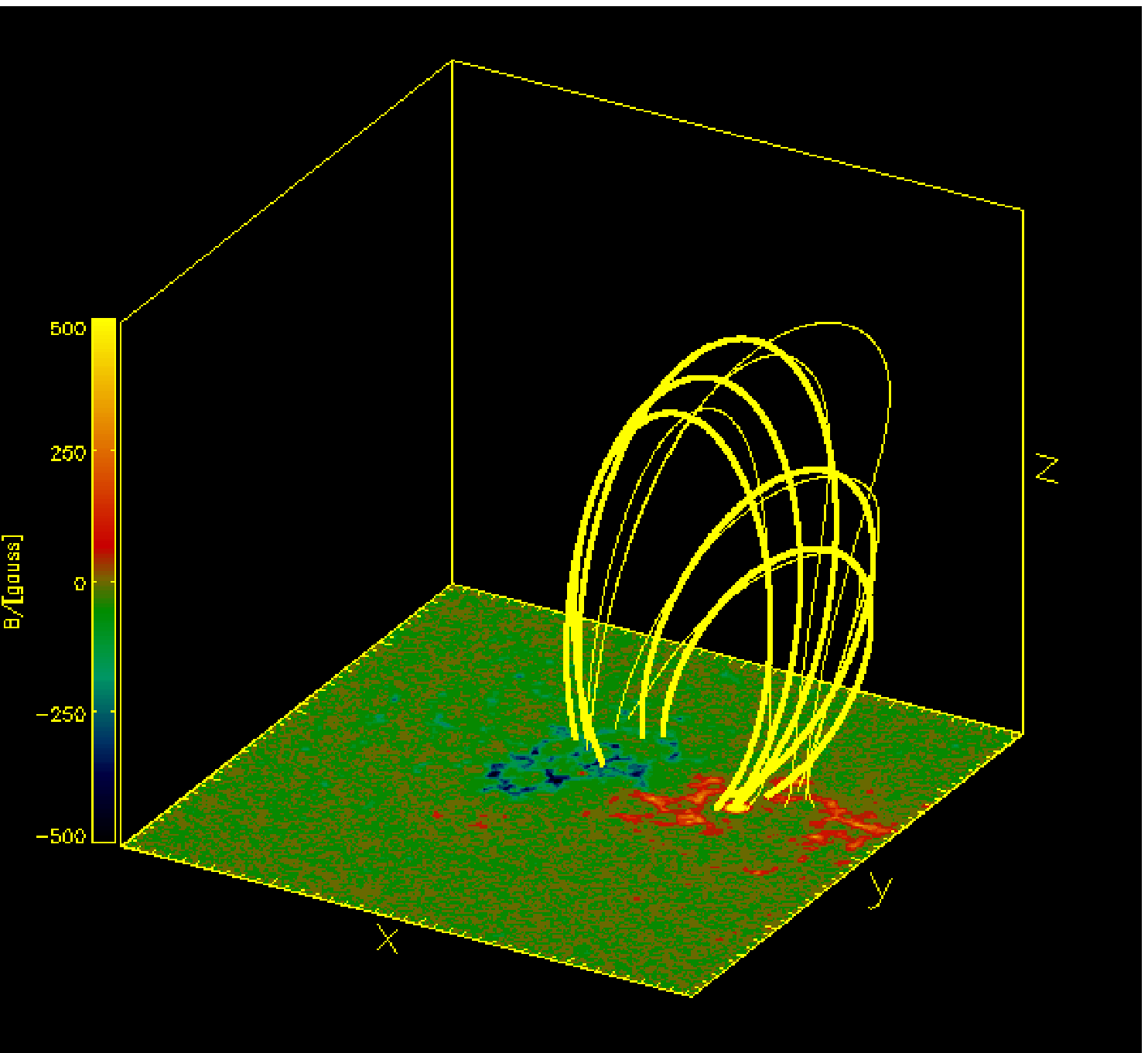}
} \caption{Upper panel: loops 1-7 and fitted magnetic field lines
for $\alpha=2.5$. Lower panel: loops 13,16,17,20,21 and fitted
magnetic field lines for $\alpha=-2.0$ } \label{groups}
\end{figure}

The results for the individual loops indicate that two basic
subgroups of loops can be identified, namely the loops 1-7
(subgroup 1) which all have positive values of $\alpha$ and the
loops $13,16,17,20,21$ (subgroup 2) which have negative values of
$\alpha$. Subgroup 2 contains only loops for which only a very
small part of the total loop has been observed so that the
circular extension is only determined by an almost straight line.
It is therefore doubtful whether the derived $\alpha$ values are
really meaningful for subgroup 2. Since we are mainly interested
in testing the method in this paper, we have nevertheless included
the results for subgroup 2, but the results are more interesting
from a methodological point-of-view. A difference between the two
subgroups is their inclination angle with the direction
perpendicular to the solar surface (see figure \ref{groups}).

In addition we notice that loop 9 is nearly potential and belongs
to neither of these subgroups. We point out that for these loops
the foot-point method and the loop distance method lead to similar
optimal values of $\alpha$ as can be seen by comparing the first
and the last two rows of Table \ref{table1}.

To assess the possible differences between fitting single loops
and groups of several loops, we minimised the sum of the
individual $C$'s over the loop subgroups,
$C(\alpha)=C_1(\alpha)+C_2(\alpha)+\dots$, $i=$ loop number, for
both subgroups identified above. As the loops in the two subgroups
are individually fitted best by $\alpha$ values with opposite
sign, it does not make sense to try and fit both groups
simultaneously. This is an obvious limitation of the method
imposed  by the linear force-free field extrapolation which can
only be overcome by non-linear force-free field extrapolation
allowing for a spatial variation of $\alpha$.

We found the optimal values of $\alpha=2.5$ for subgroup 1 and
$\alpha=-2.0$ for subgroup 2. As in the individual cases the sign
of $\alpha $ for the two loop subgroups is different. The value of
$\alpha$ for subgroup 1 is slightly lower than the average value
of the optimal $\alpha$ for that subgroup ($\bar{\alpha} \approx
2.7$). For the second subgroup, the optimal value of $\alpha$ has
to be compared to an average of the individual values of
$\bar{\alpha} \approx -1.9$.

The values of $C_i(\alpha)$ for the optimal group $\alpha$ for
each loop in a group are only slightly larger (about $5-10 \%$)
than the optimal values for single loop optimisation. We therefore
conclude that the magnetic field close to subgroup 1 can be
represented with a reasonable accuracy by a linear force-free
field with $\alpha=2.5$ and the field of subgroup 2 by
$\alpha=-2.0$. A three\--di\-men\-sio\-nal view of the loops and
the fitted field lines is shown in figure \ref{groups} where the
upper panel corresponds to subgroup 1 and the lower panel to
subgroup 2. The observed loops are indicated by thick lines and
the optimised reconstructed field lines by thin lines.

As can be clearly seen in figure \ref{groups} the main difference
between the loops in the two subgroups is that all loops in
subgroup 1 have a negative inclination angle with the vertical
whereas all loops in subgroup 2 have a positive inclination angle
(see also Table 1 in \inlinecite{aschwanden99}). As shown above
this corresponds to a positive value of $\alpha$ for subgroup 1
and a negative $\alpha$ for subgroup 2. Figure \ref{groups} also
shows that the two subgroups occupy different spatial regions
although their foot points are generally not too far apart.

We would like to emphasize again that all the results presented in
this section depend on the circular loop extrapolation used by
\inlinecite{aschwanden99}. Therefore all derived $\alpha$ values
can be considered only as accurate as the shape of the
corresponding loop used for the determination of $\alpha$. In
particular, extrapolations based only on small observed parts of
loops have to be regarded with great caution. However, we were
mainly interested in testing our method and under the assumption
that the extrapolated loops are correct, the method seems to give
consistent results.

%
\section{Conclusions and Outlook}
\label{conclusions}

In this paper we undertook a first step towards including
three\--di\-men\-sio\-nal information from stereoscopic
observations into a reconstruction of coronal magnetic fields from
photospheric magnetic field measurements. Due to the low plasma
$\beta$ in the solar corona force-free magnetic fields can be used
and in the present paper, we restricted our analysis to linear
force-free fields for simplicity. The extrapolation method uses
the line-of-sight photospheric magnetic field as a boundary
condition.

As the coronal plasma has a very high conductivity, the magnetic
field is frozen into the plasma. Therefore one can assume that the
coronal plasma structures also outline the magnetic field. This is
the fundamental assumption of our method. The idea of our
reconstruction method is to take both the photospheric magnetogram
and the three\--di\-men\-sio\-nal stereoscopic information into
account to derive the magnetic field configuration in the solar
corona. The magnetogram provides information regarding the
strength and the distribution of the magnetic fields, whereas the
three\--di\-men\-sio\-nal loop shapes restrict the current
density. The method works iteratively in that we first calculate a
magnetic field configuration from the line of sight magnetogram
for a given value of $\alpha$ without considering the stereoscopic
information. To test the configuration we compare a magnetic loop
of this magnetic field with the stereoscopic observations. This
can be done in various ways of which we have presented two
(foot-point method and loop distance method). The value of
$\alpha$ is then systematically optimised by minimising the
difference between the observed and the reconstructed loop shape.

We have applied this method to loops deduced by
\inlinecite{aschwanden99} using the method of dynamic stereoscopy
for the active region NOAA 7986 observed with SOHO/MDI and EIT on
1996 August 30. For the future we are planing to apply our method
to data taken by the SECCHI instrument aboard the STEREO mission.

The results obtained for the \inlinecite{aschwanden99} data are
promising, but indicate the need for improving the method further
by the use of non-linear force-free fields. Work along these lines
is in progress and will be presented in a forthcoming publication.

\acknowledgements
We thank Bernd Inhester, Fabrice Portier-Fozzani, Eric Priest and
Rainer Schwenn for useful discussions. We also thank the referee
for insightful comments. The data used have been provided by the
SOHO/MDI and SOHO/EIT Consortia. SOHO is a joint ESA/NASA program.
This work was supported by an EC Marie-Curie-Fellowship (TW) and a
PPARC Advanced Fellowship (TN).


\end{article}
\end{document}